\documentclass[pra,twocolumn,superscriptaddress,a4]{revtex4}

\usepackage{amsmath}
\usepackage{amssymb}
\usepackage[pdftex]{graphicx}
\usepackage{booktabs}
\usepackage{multirow}
\usepackage{hyperref}
\usepackage{color}
\usepackage[normalem]{ulem}

\def\:={\,\raisebox{0.85pt}{.}\hspace{-2.78pt}\raisebox{2.85pt}{.}\!\!=\,}
\def\=:{\,=\!\!\raisebox{0.85pt}{.}\hspace{-2.78pt}\raisebox{2.85pt}{.}\,}
\newcommand{\be}{\begin{equation}}
\newcommand{\ee}{\end{equation}}
\newcommand{\bea}{\begin{eqnarray}}
\newcommand{\eea}{\end{eqnarray}}

\begin{document}

\title{Absence of many-body localization in a continuum}
\author{I. V. Gornyi}
\affiliation{Institut f\"ur Nanotechnologie, Karlsruhe Institute of Technology, 76021 Karlsruhe, Germany}
\affiliation{\mbox{Institut f\"ur Theorie der Kondensierten Materie, Karlsruhe Institute of Technology, 76128 Karlsruhe, Germany}}
\affiliation{A.F. Ioffe Physico-Technical Institute, 194021 St.~Petersburg, Russia}
\affiliation{L.D. Landau Institute for Theoretical Physics, 119334 Moscow, Russia}
\author{A. D. Mirlin}
\affiliation{Institut f\"ur Nanotechnologie, Karlsruhe Institute of Technology, 76021 Karlsruhe, Germany}
\affiliation{\mbox{Institut f\"ur Theorie der Kondensierten Materie, Karlsruhe Institute of Technology, 76128 Karlsruhe, Germany}}
\affiliation{L.D. Landau Institute for Theoretical Physics, 119334 Moscow, Russia}
\affiliation{Petersburg Nuclear Physics Institute, 188300 St.~Petersburg, Russia}
\author{M. M\"{u}ller}
\affiliation{Condensed Matter Theory Group, Paul Scherrer Institute, CH-5232 Villigen PSI, Switzerland}
\affiliation{The Abdus Salam International Centre for Theoretical Physics, 34151, Trieste, Italy}
\email{markus.mueller@psi.ch}
\author{D. G. Polyakov}
\affiliation{Institut f\"ur Nanotechnologie, Karlsruhe Institute of Technology, 76021 Karlsruhe, Germany}

\begin{abstract}
{We show that many-body localization, which exists in tight-binding models, is unstable in a continuum. Irrespective of the dimensionality of the system, many-body localization does not survive the unbounded growth of the single-particle localization length with increasing energy that is characteristic of the continuum limit. The system remains delocalized down to arbitrarily small temperature $T$, although its dynamics slows down as $T$ decreases. Remarkably, the conductivity vanishes with decreasing $T$ faster than in the Arrhenius law. The system can be characterized by an effective $T$-dependent single-particle mobility edge which diverges in the limit of $T\to 0$. Delocalization is driven by interactions between hot electrons above the mobility edge and the ``bath" of thermal electrons in the vicinity of the Fermi level.}
\end{abstract}

\maketitle

\section{Introduction}
\label{s1}

The profound effect that disorder can have on quantum particles, known as Anderson localization \cite{anderson58}, persists when the particles interact with each other. One of the most exciting aspects of the physics that emerged in this field concerns the influence of electron-electron interactions on Anderson localization at nonzero temperature $T$, in the absence of coupling between the electron system and external degrees of freedom. Specifically, the realization \cite{fleishman80,gornyi05,basko06} that localization may survive in an interacting system at nonzero $T$ started a major trend in the physics of disordered many-body systems, with important implications for general statistical mechanics, and the subject, running under the name of many-body localization, continues to develop at a rapid pace \cite{nandkishore15}.

It is by now commonly accepted wisdom that disordered interacting models, either fermionic or spin models, that are formulated {\it on a lattice} generically exhibit many-body localization. Prime examples, also intensively studied numerically, are the tight-binding models of lattice fermions in one dimension and the related spin chain models. In particular, a proof of the existence of many-body localization with emphasis on mathematical rigor was proposed in Ref.~\cite{imbrie16} for the case of a random Ising chain in a transverse field. One of the frequently considered limits for the lattice models is the limit of an infinitely large $T$ \cite{oganesyan07}, which means that the results do not essentially change with varying $T$ when $T$ is much larger than the bandwidth.

On the other hand, in Refs.~\cite{gornyi05} and \cite{basko06}, the analytical results for many-body localization were obtained for disordered interacting electron systems {\it in a continuum} (no lattice), i.e., with no upper bound on the bandwidth of the electron dispersion relation (this is true both for the model with a weak random potential from Ref.~\cite{gornyi05} and the ``granular model" from Ref.~\cite{basko06}). Both papers \cite{gornyi05,basko06} arrived at the conclusion that there is a finite-$T$ localization-delocalization transition. Delocalization in real space was understood in Ref.~\cite{gornyi05} as being related to delocalization on a certain graph in Fock space \cite{altshuler97}, with the insulating phase being destabilized by ``ballistic" paths (in both spaces) with no self-intersections. The analysis of the delocalizing effect of various classes of paths in Fock space (including the upper bound for delocalization from the ballistic paths) was corroborated in detail in Ref.~\cite{ros15}. In Ref.~\cite{basko06}, the stability of the localized and delocalized phases was analyzed by means of a self-consistent treatment of the perturbation series. The structure of the perturbative expansion for the hybridization spreading in Fock space, within the framework of Refs.~\cite{gornyi05} and \cite{basko06}, was recently discussed in Ref.~\cite{gornyi16}, with emphasis on a closely related quantum-dot problem from Ref.~\cite{altshuler97}.

More recently, it was realized \cite{gornyi16a} that Refs.~\cite{gornyi05,basko06,ros15}, and \cite{gornyi16} missed an important ingredient of the theory, namely the effect of spectral diffusion in the hybridization process, that greatly facilitates ``many-body delocalization." More precisely, spectral diffusion pulls down the upper bound for delocalization by a parametrically large factor for the case \cite{gornyi05,basko06,ros15,gornyi16} of weak interaction. All in all, however, the conventional wisdom maintains that the models defined in a continuum, similar to the models on a lattice, have the many-body localized phase, i.e., are characterized by a critical temperature below which the system is localized in dimensions $D=1$ and 2, even though this was questioned based on the consideration of anomalously rare high-energy processes \cite{ros15,deroeck16,kagan85}. The similarity in this regard between the lattice and continuum models seems especially reasonable when viewed from the commonly accepted perspective that delocalization is mediated, in the first place, by electron-electron scattering between states within an energy band of width $T$ around the Fermi surface. Within this perspective, one can think of an effective bandwidth in the continuum model being given by $T$ (with possible interaction-induced renormalization effects, which come from larger energy scales, only renormalizing the parameters of the system within the effective bandwidth).

In this paper, we address the question of whether genuine many-body localization indeed exists in a continuum by relaxing the condition, assumed in the effective low-energy models \cite{gornyi05,basko06}, that the electron system is characterized by an energy independent single-particle localization length. The rationale behind this question is that the single-particle localization length $\xi(\epsilon)$ for electrons moving at energy $\epsilon$ in a continuum in the presence of a random potential, for $D=1$ and 2, generically grows without bound as $\epsilon$ is increased. That is, the question is about a competition, in the context of many-body delocalization, between the growth of $\xi(\epsilon)$ and the falloff of the thermal distribution function $f(\epsilon)=1/[e^{(\epsilon-\epsilon_F)/T}+1]$ with increasing $\epsilon$.

Earlier, the above question was posed in Ref.~\cite{nandkishore14} for a weakly disordered two-dimensional system in a continuum (in the case of time-reversal symmetry), for which $\xi(\epsilon)$ is an exponential function of $\epsilon$, similar to $f(\epsilon)$. It was understood that high-energy single-particle states can be important for many-body delocalization, despite being sparsely populated in the tail of the distribution function. The conclusion in Ref.~\cite{nandkishore14} was that there still is (up to the ``mobile hot-bubble scenario" \cite{ros15,deroeck16,kagan85}) a localization-delocalization transition in two dimensions at a critical temperature (given by the disorder-induced single-particle scattering rate, i.e., independent of the interaction strength), even though $\xi(\epsilon)$ diverges so rapidly with increasing $\epsilon$. Importantly, it was realized in Ref.~\cite{nandkishore14} that the divergency of $\xi(\epsilon)$ as $\epsilon$ increases can strongly enhance many-body delocalization by parametrically lowering the critical temperature, although the enhancement occurs in Ref.~\cite{nandkishore14} only in the limit of a small strength of interaction. For one dimension, Ref.~\cite{nandkishore14} argued the growth of $\xi(\epsilon)$ with increasing $\epsilon$ in a continuum to be irrelevant to the stability of many-body localization.

The picture of many-body delocalization from Ref.~\cite{nandkishore14} is in contrast to the one proposed in Ref.~\cite{mueller09} for a closely related problem in which there exists a zero-$T$ mobility edge and $\xi(\epsilon)$ diverges when approaching the edge from below. It was argued in Ref.~\cite{mueller09} that inelastic collisions at finite (but arbitrarily small) $T$ delocalize excitations with sufficiently large $\xi(\epsilon)$, since the inverse lifetime of these excitations exceeds the relevant (to the Thouless argument) level spacing. As a result, there emerges an effective $T$-dependent mobility edge (lower than the $T=0$ edge) which governs ``superactivated" transport (faster suppression with decreasing $T$ than in Arrhenius' law). When applied to the situation in which $\xi(\epsilon)$ diverges at $\epsilon\to\infty$, as in Ref.~\cite{nandkishore14}, the reasoning presented in Ref.~\cite{mueller09} would eliminate the possibility of a finite-$T$ transition.

Below, we derive a few basic results in this direction and argue that many-body localization does not survive the unbounded growth of $\xi(\epsilon)$ with increasing $\epsilon$, in contrast to Ref.~\cite{nandkishore14}. The true localization-delocalization transition at a critical temperature, commonly assumed for $\xi(\epsilon)={\rm const}$, transforms into a crossover, so that the system in a continuum remains delocalized down to arbitrarily small $T$, albeit with progressively slower dynamics as $T$ is decreased. In the limit of low $T$, we predict superactivated transport of charge and energy.

Perhaps most remarkably, our prediction holds true irrespective of dimensionality. More precisely, we find, for the lowest-order channel of delocalization, that one-dimensional systems in a continuum only remain ``marginally" stable against delocalization in the limit of low $T$ for the model case of exactly zero-range disorder. Otherwise, the breakdown of localization occurs in one dimension as well, similarly to systems of higher dimensionality that are delocalized for an arbitrary type of disorder.

\section{Many-body delocalization in the hot-cold mixture}
\label{s2}

\begin{figure}
\centerline{\includegraphics[width=\columnwidth]{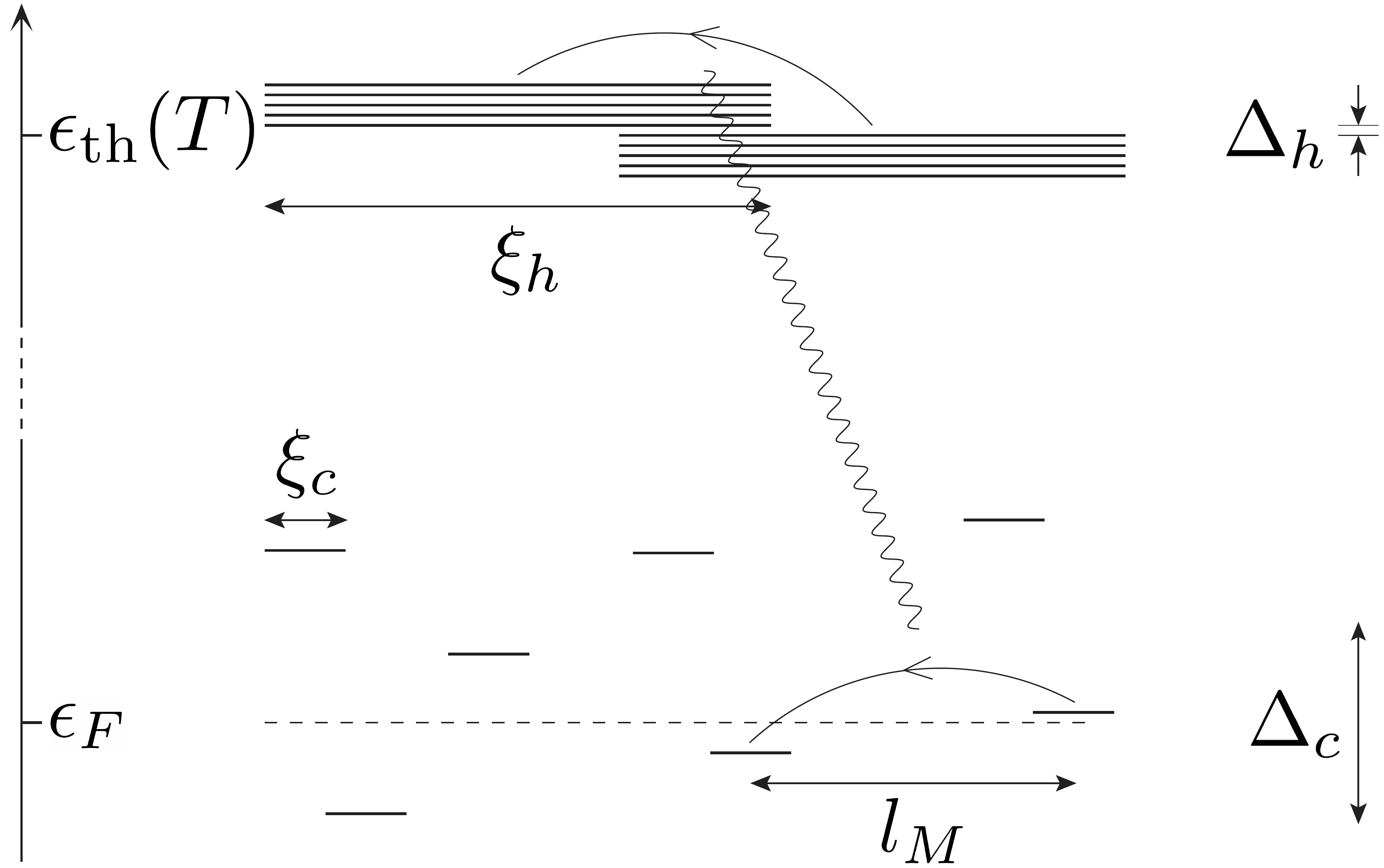}}
\caption{Schematic picture of resonant coupling between electron states with vastly different single-particle localization lengths, $\xi_c$ at the Fermi level $\epsilon_F$ and $\xi_h\gg\xi_c$ at the threshold energy $\epsilon_{\rm th}(T)$, with the corresponding single-particle mean level spacings in the localization volume, $\Delta_c$ and $\Delta_h\ll\Delta_c$, respectively. Each of the two subsystems of ``hot" and ``cold" electrons has a width in energy space of the order of the temperature $T$, with $\Delta_h\ll T\ll\Delta_c$ in the limit of low $T$. Typical resonant processes correspond to hopping distances in the cold sub-system given by $l_M\simeq 2\xi_c\ln(\Delta_c/T)$.}
\label{f1}
\end{figure}

Consider a disordered interacting system in a continuum that would be many-body localized if $\xi(\epsilon)$ was independent of $\epsilon$. In our approach, highly excited electron states, with $\epsilon$ larger than a certain $T$-dependent threshold energy $\epsilon_{\rm th}(T)$ to be determined below [the critical energy is well defined in the limit $\epsilon_{\rm th}(T)-\epsilon_F\gg T$], are delocalized by interactions with ``thermal" electrons (those with $|\epsilon-\epsilon_F|\sim T$). In turn, delocalization extends over the whole energy spectrum, but the very possibility of ruining the stability of many-body localization derives from the hybridization of highly excited electron states among themselves by means of interactions with thermal electrons. The delocalization at $\epsilon >\epsilon_{\rm th}(T)$ relies on the exchange of energy but not charge between the highly excited and thermal electrons, with electron states with intermediate energies playing no role in it. That is, one can think of the system of ``hot" electrons and the bath of ``cold" electrons as of two subsystems each of which has a width in energy space of the order of $T$, as illustrated in Fig.~\ref{f1}. Below, the quantities that refer to the two subsystems of hot and cold electrons, such as the localization length $\xi$, the density of states $\rho$, the mean level spacing in the localization volume $\Delta=1/\rho\xi^D$, etc., are labeled with $h$ and $c$, respectively.

Within our approach, the parameter that controls many-body delocalization of the $h$-system is
\be
\eta_{hc}=V_{hc}/\Delta_{3hc}~,
\label{9}
\ee
where $V_{hc}$ is the characteristic amplitude of the matrix element of interaction for coupling between the $h$- and $c$-systems and $\Delta_{3hc}$ is the characteristic level spacing of final three-particle states for the decay of an $h$-state into another $h$-state and a particle-hole pair of $c$-states. This parameter is generically $\epsilon$ dependent and may possibly grow as $\epsilon$ increases. If $\eta_{hc}$ exceeds a certain critical value (which is model dependent but generically of the order of unity), there emerges an infinite resonant network of couplings between $h$-states, mediated by interactions with $c$-states. This only becomes possible because of a finite density of thermal excitations in the $c$-system to support the hybridization of an $h$-state by alternately losing and acquiring energy of the order of $T$. This is the reason why the critical energy $\epsilon_{\rm th}(T)$ mentioned above is a function of $T$. It is important here that, although there are many more resonant couplings available for $h$-electrons at every step of the hybridization process, namely those with energy transfers much larger than $T$, the infinite expansion of the resonant network is bottlenecked by those resonant transitions in which an $h$-electron absorbs energy supplied by $c$-electrons.

Crucially, the parameter (\ref{9}) does not contain the occupation number $f(\epsilon)$ for the propagating $h$-state. Indeed, the hybridization between two $h$-states accompanied by the excitation of an electron-hole pair in the $c$-system does not require that each of the $h$-states be occupied. This is because, once the initial $h$-state is occupied, it can hybridize with other $h$-states by means of electron-electron interactions without ever receiving energy from within the $h$-system. As a consequence, the delocalization process within the $h$-system proceeds much more efficiently through interactions between the $h$- and $c$-electrons rather than between $h$-electrons only.

The above picture is in stark contrast to Ref.~\cite{nandkishore14}, where the parameter that controls many-body delocalization was argued to be proportional to the occupation number of the {\it initial} single-particle state [see, e.g., Eqs.~(11) or (A3) there], being then exponentially small in $\epsilon/T$---even for the case of interactions between the $h$- and $c$-systems. Moreover, while the parameter $\eta_{hc}$ from Eq.~(\ref{9}) describes how a single $h$-state decays and is, therefore, proportional to the total number of available final states, the parameter in Ref.~\cite{nandkishore14} also contains the total number of {\it initial} states (in the localization volume) that can possibly decay. We cannot justify the insertion of the occupation number of the initial single-particle state in the stability criterion for many-body localization. The difference in the criteria of many-body delocalization affects both the case of interactions between the $h$- and $c$-systems and the case of interactions present within the $h$-system only (both discussed in Ref.~\cite{nandkishore14}, with emphasis on the latter case, which gives, in Ref.~\cite{nandkishore14}, a lower critical temperature for the localization-delocalization transition in two dimensions).

It is worth mentioning that our primary focus here is on the demonstration of many-body delocalization persisting, in a continuum, to arbitrarily low $T$ and not on the analysis of relative contributions to it of various alternative hybridization channels. In this paper, we treat weak interactions at the lowest order, namely we deal with resonant couplings between sites in Fock space of first order in the interaction strength. If these destroy many-body localization (which is the case, as shown below), we consider genuine many-body localization as nonexistent. 
Instead of being strictly absent, low-$T$ transport is dramatically suppressed, but remains finite at any nonzero $T$. It is important here that, in the limit of weak interaction, the different hybridization channels that correspond to the resonant couplings of different orders in the interaction strength are likely to be additive as far as their action on the stability of the localized phase is concerned. We cannot exclude that higher-order interactions are even more effective in destabilizing many-body localization. One of the additional mechanisms of delocalization which arises when expanding the resonant couplings to second order in the interaction strength was discussed in a closely related context in Ref.~\cite{nandkishore14a}. Another possible channel of delocalization can involve resonant hybridization of more complex excitations that include multiple hot electrons, see discussion in Sec.~\ref{s5}.

\section{One dimension}
\label{s3}

It is instructive to start by considering the ``worst" case from the point of view of the possibility of delocalization in the limit of low $T$, namely the case of a slow (power-law) growth of $\xi(\epsilon)$ with increasing $\epsilon$ in one dimension. Assume that the random potential $U(x)$ is weak everywhere (Gaussian correlated) and short-ranged, with the correlator $\langle U(x)U(0)\rangle=w\delta(x)$ (``white noise"). Then, for a quadratic dispersion relation $\epsilon=k^2/2m$, where $k$ is the electron wavevector, $\xi(\epsilon)$ is a linear function of $\epsilon$:
\be
\xi(\epsilon)=\frac{2}{mw}\,\epsilon~.
\label{1}
\ee
We defined here, for definiteness, $\xi(\epsilon)$ as the backscattering mean free path. The white-noise model for the correlations of $U(x)$ provides the slowest possible growth of $\xi(\epsilon)$ with increasing $\epsilon$ in a continuum. In particular, if disorder is characterized by a single correlation radius $d$, then $\xi(\epsilon)$ changes with $\epsilon$ faster than linearly for $|k|d\agt 1$. Note also that, in one dimension, the dependence of $\xi(\epsilon)$ on $\epsilon$ for the limit of short-range disorder is entirely due to the nonlinearity of the dispersion relation, i.e., due to the $\epsilon$ dependence of the velocity $v(\epsilon)$ and, correspondingly, of the density of states (per spin) $\rho(\epsilon)=1/\pi v(\epsilon)$.

Let us focus on the (least favorable to delocalization) limit of low $T\ll\Delta_c$. The (effective) spatial density of thermal excitations in the $c$-system for $T\ll\Delta_c$ can be conveniently defined as
\be
n_T=2\!\int\!\!\!\!\!\!-\,dx\int_{-\infty}^\infty\!\!d\epsilon\int_{-\infty}^{0-}\!\!d\omega\,f(\epsilon)\,[\,1-f(\epsilon+\omega)\,]\,[-S_\omega(x)]
\label{2}
\ee
in terms of the spectral function
\begin{multline}
S_\omega(x)\\
=\left\langle\sum_{ij}\delta(\epsilon_i-\epsilon)\delta(\epsilon_j-\epsilon-\omega)\psi_i(0)\psi_j(0)\psi_i(x)\psi_j(x)\right\rangle~,
\label{3}
\end{multline}
written here as the disorder-averaged sum over the exact single-particle states $\psi_i$ with energies $\epsilon_i$. The dash in $\int\hspace{-3.1mm}-$ means that the integral is taken over $|x|\gg\xi_c$ (the reason for introducing this constraint and the factor of 2 in the definition of $n_T$ will become clear shortly). For $|\omega|\ll\Delta_c$ and $|x|\gg\xi_c$, $S_\omega(x)$ is given (in the vicinity of its maximum) by \cite{gorkov83}
\be
S_\omega(x)\simeq -\frac{1}{2}\,\rho_c^2\left(\frac{\xi_c}{\pi l_M}\right)^{1/2}\exp\left[\,-\frac{(|x|-l_M)^2}{4\xi_c\,l_M}\,\right]~,
\label{4}
\ee
where $l_M=2\xi_c\ln (\Delta_c/|\omega|)\gg \xi_c$ is the Mott length \cite{mott70}. In the low-frequency limit, $S_\omega(x)$ is sharply peaked at $|x|=l_M$ because of the level repulsion for smaller distances and the lack of resonant hybridization for larger. The minus sign in Eq.~(\ref{4}) reflects the fact that the resonance at frequency $|\omega|\ll\Delta_c$ occurs between the symmetric and antisymmetric combinations of two spatially separated (by the distance $l_M$) spikes in the hybridized wavefunctions. This means that $S_\omega(x)$, apart from being peaked at $|x|=l_M$, has one more peak---of the opposite sign---centered at $x=0$ (this fulfils the general requirement that $\int\!dx\,S_\omega(x)=0$ for $\omega\neq 0$). The peak at $x=0$ has a width of the order of $\xi_c$, hence the condition $|x|\gg\xi_c$ and the factor of 2 in $n_T$.

We thus have, for $T\ll\Delta_c$,
\be
n_T\simeq\frac{2\pi^2}{3}\,\frac{1}{\xi_c}\left(\frac{T}{\Delta_c}\right)^2~.
\label{5}
\ee
The factor $(T/\Delta_c)^2$ comes from the constraint that the energies of both the initial and final states [$\epsilon_i$ and $\epsilon_j$ in Eq.~(\ref{3})] be within the energy band of the characteristic width $T$ around the Fermi level. It is worth noting that the Mott length (with $|\omega|\sim T$),
\be
l_M\to 2\xi_c\ln (\Delta_c/T)~,
\label{6}
\ee
which gives the ``hopping distance" for transitions in Eq.~(\ref{2}), does not show up in Eq.~(\ref{5}) (this is a peculiarity of one dimension, see below).

The product
\be
N_{\rm eff}=n_T\xi_h\sim (\xi_h/\xi_c)(T/\Delta_c)^2\propto\epsilon
\label{7}
\ee
gives, for $N_{\rm eff}\agt 1$, the characteristic number of electron-hole pairs in the $c$-system that can potentially  hybridize with a typical state in the $h$-system by exchanging energy of the order of $T$. Recall that it is the resonant transitions that absorb energy from the $c$-system that are a bottleneck in the expansion of the resonant network to infinity. This is why we can think of the effective coordination number for resonant couplings between sites in Fock space as given by Eq.~(\ref{7}), although there are many more resonant couplings available for an $h$-electron at every step in the hybridization process, with larger energy transfers. The characteristic level spacing $\Delta_{3hc}$ of final three-particle states reads, then, as
\be
\Delta_{3hc}=\frac{1}{N_{\rm eff}}\,\Delta_h\propto \frac{1}{\epsilon^{3/2}}~.
\label{8}
\ee
The vanishing of $\Delta_{3hc}$ as $\epsilon$ is increased is the reason why the $h$-system may potentially destabilize many-body localization in the limit of low $T$. Whether this indeed occurs depends on the behavior, with increasing $\epsilon$, of the parameter $\eta_{hc}$ [Eq.~(\ref{9})].

We now turn to the question of how $V_{hc}$ changes as $\epsilon$ is increased. We work with the assumption that the two-body interaction $V(x)$ is short-ranged, namely its radius $a$ is much smaller than any relevant spatial scale for direct (Hartree) interactions between $h$- and $c$-electrons (recall that the characteristic energy transfer for these interactions is given by $T$), i.e., $a\ll\min\{\xi_{c,h}\,,\,v_{c,h}/T\}$. For $T\ll\Delta_c$, as assumed above, the strongest condition is $a\ll\xi_c$. We focus, for definiteness, on spinless electrons, for which $a$ cannot be sent to zero, because otherwise there is an exact cancellation between the Hartree and exchange interactions.

The matrix element for coupling between two $h$-states $i$ and $j$ and two $c$-states $k$ and $l$ is written as
\bea
V_{ijkl}&\!=\!&\int\!dx_1\!\int\!dx_2\,\psi^h_i(x_1)\psi^c_k(x_2)V(x_1-x_2)\nonumber\\
&\!\times\!&\left[\,\psi^h_j(x_1)\psi^c_l(x_2)-\psi^c_l(x_1)\psi^h_j(x_2)\,\right]~,
\label{10}
\eea
where the first and second terms in the square brackets describe direct and exchange interactions, respectively. The matrix element $V_{hc}$ in Eq.~(\ref{9}) can be defined as the disorder-averaged root mean square of $V_{ijkl}$ from Eq.~(\ref{10}). Let us first look at the Hartree term.

Each of the wavefunctions in Eq.~(\ref{10}) is a rapidly oscillating function of $x$ with a smoothly varying envelope: in the $h$-system, the wavevector of the oscillations is given by $2\epsilon/v_h$ and the characteristic spatial scale for the smooth variation is given by $\xi_h$, and similarly for the $c$-system. Because of the separation of scales $l_M\ll\xi_h$ and $l_M\ll v_h/T$ [with $l_M$ from Eq.~(\ref{6})], we can exploit the property of the $h$- and $c$-states that the product $\psi^h_i\psi^h_j$ (apart from the part rapidly oscillating with the wavevector $4\epsilon/v_h$) is a smooth function on the spatial scales provided by the product $\psi^c_k\psi^c_l$. In the first approximation, removing $\psi^h_i\psi^h_j$ from under the integral sign in Eq.~(\ref{10}) (with the intention of substituting $1/\xi_h$, by an order of magnitude, for it) gives exactly zero for the Hartree term [for arbitrary $V(x)$]. That is, the leading contribution to the Hartree part $V_{hc}^{\rm H}$ of $V_{hc}$ is obtained by expanding $\psi^h_i(x_1)\psi^h_j(x_1)$ to linear order in $x_1$ around the position of the $c$-states (which produces a dipole matrix element $\int\!dx\,\psi^c_kx\psi^c_l$ for scattering between $\psi^c_k$ and $\psi^c_l$). In the limit of large $\epsilon$ (namely for $\Delta_h \ll T$), the main term in the expansion in $x_1$ comes from the beating between the oscillatory factors in $\psi^h_i(x_1)$ and $\psi^h_j(x_1)$ (and not from the envelopes of the wavefunctions), which gives
\be
V^{\rm H}_{hc}\sim\alpha T\,\frac{l_M}{\xi_h}\propto\frac{1}{\epsilon^{3/2}}~,
\label{11}
\ee
where the ``coupling constant"
\be
\alpha=\rho_h|u|\propto\frac{1}{\epsilon^{1/2}}~,\qquad u=\!\int\!dx\,V(x)~.
\label{12}
\ee
This is the contribution to the Hartree term which one obtains by sending $a$ to zero, i.e., for $V(x)=u\delta(x)$, and which is, therefore, cancelled by its exchange counterpart in this limit.

The important point is that $V^{\rm H}_{hc}$ from Eq.~(\ref{11}) falls off with increasing $\epsilon$ as $1/\epsilon^{3/2}$, i.e., in precisely the same manner as $\Delta_{3hc}$ does [Eq.~(\ref{8})]. It follows that (weak) direct interactions cannot possibly provide an energy scale above which the parameter $\eta_{hc}$ [Eq.~(\ref{9})] is large enough to destabilize many-body localization for the case of white-noise disorder. 
This is because in the limit of large $\epsilon$ any further terms in the expansion of $V^{\rm H}_{hc}$ in $a$ vanish faster than $V^{\rm H}_{hc}$ in Eq.~(\ref{11}).

Now, let us turn to the exchange term in Eq.~(\ref{10}). Here, the situation in the hot-cold mixture is different in that taking $\psi^h_i\psi^h_j$ out from under the integral sign in Eq.~(\ref{10}) does not make the integral vanish (provided $a$ is not zero), in contrast to the Hartree term. The thus obtained contribution to the characteristic amplitude $V^{\rm ex}_{hc}$ of the exchange part of the matrix element (\ref{10}) depends on $\epsilon$ only through the factor $1/\xi_h$ (which comes from the smooth part of the product $\psi^h_i\psi^h_j$ taken at the position of the $c$-states). More precisely, it scales with $\epsilon$ and $a$ in the limit of small $a$ as $ua^2/\epsilon$, with a prefactor solely determined by the properties of the $c$-system. This term in $V^{\rm ex}_{hc}$ falls off with increasing $\epsilon$ more slowly than $\Delta_{3hc}$. However, this does not mean that the main contribution to $V_{hc}$ in the limit of large $\epsilon$ comes from exchange interactions. This is because the energy transfer in the exchange term is given by $\epsilon-\epsilon_F$ (in contrast to $T$ in the Hartree term), so that pulling $\psi^h_i\psi^h_j$ out from under the integral sign in the exchange term is only legitimate for $\epsilon\ll 1/ma^2$. In the limit of larger $\epsilon$, the effective coupling constant for exchange interactions is given by $\rho_h|\!\int\!dx\,V(x)\cos(2\epsilon x/v_h)|\ll\alpha$, whereas it is still given by $\alpha$ from Eq.~(\ref{12}) for direct interactions. It follows that $V_{hc}$ in the limit of large $\epsilon$ is determined by $V^{\rm H}_{hc}$, i.e., vanishes with increasing $\epsilon$ as $1/\epsilon^{3/2}$.

The fact that $V_{hc}$ and $\Delta_{3hc}$ in the limit of large $\epsilon$ scale with $\epsilon$ in precisely the same way (as $1/\epsilon^{3/2}$) means that the low-$T$ many-body localized phase in the one-dimensional system with white-noise disorder survives, at least for weak interactions, the unbounded growth of $\xi(\epsilon)$. 
Recall, however, that we have restricted ourselves here to resonant couplings of the lowest order in the interaction strength, see also the very end of Sec.~\ref{s2}. On the other hand, after we have selected the resonant subnetwork that consists of the $h$- and $c$-systems, the problem of many-body localization in Fock space bears a close resemblance to that of  single-particle localization on a Bethe lattice with connectivity $N_{\rm eff}$~\cite{abouchacra73, altshuler97}. Using the results for the Bethe lattice, the delocalization criterion would be relaxed to $\eta_{hc}\sim 1/\ln N_{\rm eff}$, because of resonances at larger distances in Fock space that are mediated by higher-order couplings. This criterion would be satisfied by sufficiently large $\epsilon$, even in the case of white-noise disorder for $D=1$. It is worth noting, however, that while the reduction of the total resonant network to the bottlenecking part of it is sufficient for the analysis of transport due to the lowest-order resonant couplings, the statistics of higher-order resonances within the $h$-system may be modified by the presence of resonances with higher energy transfers. This issue is beyond the scope of the present paper.

The marginal character of the stability of localization in the model of white-noise disorder, with this model just shown to be on the very verge of delocalization, is also a clear indication for the absence of many-body localization in any one-dimensional system in which the correlation radius of disorder $d$ is nonzero. Indeed, while (the Hartree contribution to) $V_{hc}$ [Eq.~(\ref{11})] scales with $\xi_h$ as $1/\xi_h$, the three-particle spacing of final states $\Delta_{3hc}$ [Eqs.~(\ref{7}) and (\ref{8})] scales with $\xi_h$ faster, namely as $1/\xi_h^2$. As a consequence, $\eta_{hc}$ [Eq.~(\ref{9})] grows without bound as $\epsilon$ increases for arbitrary functions $\xi(\epsilon)$ that are faster than linear. As already mentioned below Eq.~(\ref{1}), $\xi(\epsilon)\propto\epsilon$ is the slowest possible divergency of $\xi(\epsilon)$ for $\epsilon\to\infty$, changing generically to a faster growth for $|k|d\agt 1$. From this vantage point, genuine many-body localization is absent in a continuum even in one dimension, with the disclaimer ``for arbitrarily small but finite $d$."

For an arbitrary $\epsilon$ dependence of $\xi_h$, the parameter $\eta_{hc}$ is written for $T\ll\Delta_c$ as
\be
\eta_{hc}\sim\alpha\,\frac{\Delta_c}{\Delta_h}\!\left(\frac{T}{\Delta_c}\right)^3\ln\frac{\Delta_c}{T}~.
\label{13}
\ee
The product of the first two factors $\alpha\Delta_c/\Delta_h$ is a growing function of $\epsilon$ in the limit of large $\epsilon$ for nonzero $d$, which guarantees the existence of the energy $\epsilon_{\rm th}(T)$, introduced in Sec.~\ref{s2}, for which $\eta_{hc}\sim 1$. The quantity $\epsilon_{\rm th}(T)$ is model dependent, being determined by the large-$\epsilon$ asymptotic of the Fourier transform of the correlator of the random potential $\int\!dx\,\langle U(x)U(0)\rangle\cos (4\epsilon x/v)$, but has the universal property that it grows to infinity with lowering $T$. Consider two representative examples in which $\langle U(x)U(0)\rangle$ is given by $wd/[\pi(x^2+d^2)]$ and $(w/2d)\exp (-|x|/d)$. In the former case, we have
\be
\epsilon_{\rm th}(T)\simeq\epsilon_0\ln^2\chi(T)~,
\label{13a}
\ee
in the latter:
\be
\epsilon_{\rm th}(T)\sim\epsilon_0\chi(T)~,
\label{13b}
\ee
where $\epsilon_0=1/8md^2$ is the characteristic energy above which $\xi(\epsilon)$ changes the linear dependence on $\epsilon$ to a faster one and
\be
\chi(T)=\frac{w}{|u|\Delta_c}\left(\frac{\Delta_c}{T}\right)^3\frac{1}{\ln (\Delta_c/T)}\gg 1~.
\label{13c}
\ee
In both cases, $\epsilon_{\rm th}(T)$ grows with decreasing $T$ (as well as with decreasing strength of interaction and with increasing strength of disorder).

The energy $\epsilon_{\rm th}(T)$ defines an effective $T$-dependent mobility edge (cf.\ Ref.~\cite{mueller09}), such that there emerges a ``conduction band" for $\epsilon>\epsilon_{\rm th}$ populated by delocalized excitations with the effective spatial density
\be
n_h=\frac{T}{\pi[\,2\epsilon_{\rm th}(T)/m\,]^{1/2}}\,\exp\left[\,-\frac{\epsilon_{\rm th}(T)-\epsilon_F}{T}\,\right]~,
\label{14}
\ee
where the exponential factor is the occupation probability $f(\epsilon)$ of the $h$-states with $\epsilon=\epsilon_{\rm th}(T)\gg T$. It is worth emphasizing that this is the only place in our approach where $f(\epsilon)$ for the $h$-system shows up, with the parameter $\eta_{hc}$ being independent of it.

One of the important aspects of our way of thinking about many-body localization in the mixture of electrons with vastly different single-particle localization lengths is that we have, in effect, a coupled system of two electronic {\it baths}, the $c$-system and the $h$-system. Recall that, in the above, the $c$-system serves, in the first place, as a bath of localized excitations, with the spatial density $n_T$ [Eq.~(\ref{5})]. These mediate the hybridization process within the initially localized $h$-system, which leads to the creation of delocalized excitations, with the spatial density $n_h$ [Eq.~(\ref{14})]. The point to notice is that many-body delocalization occurs self-consistently in both the $h$- and $c$-systems. In effect, the $h$-system provides a bath of delocalized excitations for lower-lying electron states, so that many-body delocalization extends to the entire energy spectrum, also below $\epsilon_{\rm th}(T)$.

\begin{figure}
\centerline{\includegraphics[width=\columnwidth]{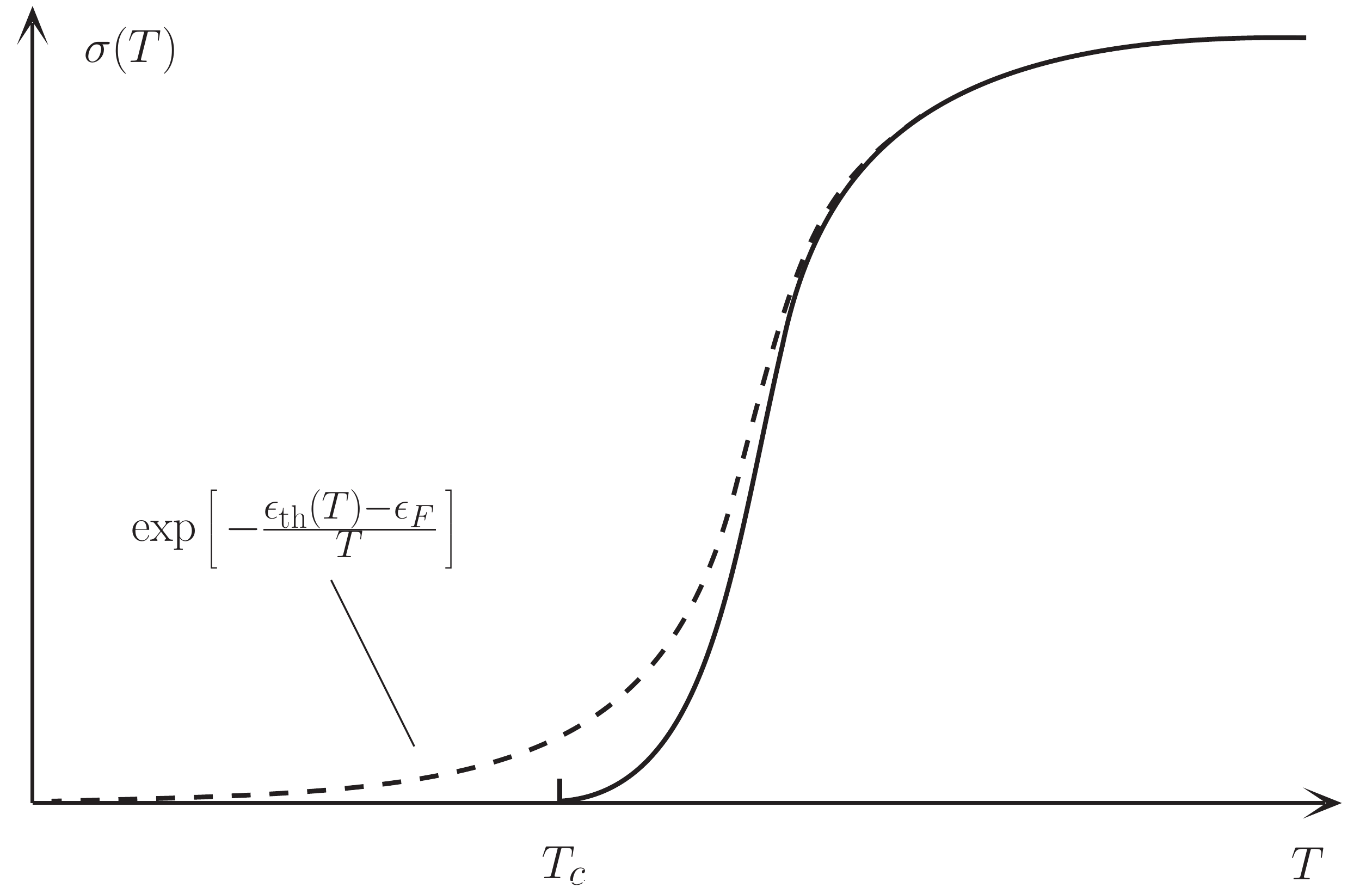}}
\caption{Schematic behavior of the conductivity $\sigma$ as a function of $T$ in a disordered interacting electron system with the single particle localization length $\xi(\epsilon)$ independent of the electron energy $\epsilon$ (solid line) and with $\xi(\epsilon)$ growing without bound as $\epsilon$ increases (dashed line), as is the case in a continuum. The localization-delocalization transition at the critical temperature $T_c$ is smeared out into a crossover by the unbounded growth of $\xi(\epsilon)$. In a continuum, $\sigma(T)$ vanishes with lowering $T$ faster than in the Arrhenius law, with the effective mobility edge $\epsilon_{\rm th}(T)$ increasing as $T$ decreases. }
\label{f2}
\end{figure}

From this perspective, the significance of the energy $\epsilon_{\rm th}(T)$ is that transport (of both charge and energy) in the coupled system of $h$- and $c$-electrons is bottlenecked by the amount of excitations above $\epsilon_{\rm th}(T)$. Specifically, in the limit of low $T$, the charge conductivity $\sigma (T)$ in the hot-cold mixture is proportional to $n_h$, i.e., obeys [with a proper normalization of $\sigma(T)$]
\be
\ln\sigma(T)\simeq -\frac{\epsilon_{\rm th}(T)-\epsilon_F}{T}~.
\label{15}
\ee
Note that, unlike in the case of an ungapped (phononlike) bath, the leading $T$-dependence from Eq.~(\ref{15}) is characteristic of the contributions to $\sigma(T)$ of both activation to the effective mobility edge at $\epsilon_{\rm th}(T)$ and variable-range hopping between electron states with energies below $\epsilon_{\rm th}(T)$. The difference between the two  manifests itself only in the subleading terms in the right-hand side of Eq.~(\ref{15}). Most unconventionally, however, transport in the hot-cold mixture is seen to be ``superactivated," with $\sigma(T)$ vanishing in the limit of low $T$ faster than in the Arrhenius law. The behavior of $\sigma(T)$ with varying $T$ is schematically illustrated in Fig.~\ref{f2}, with the localization-delocalization transition smeared out by the unbounded growth of $\xi(\epsilon)$.

Another peculiar feature of many-body delocalization in the hot-cold mixture concerns the nature of charge or energy transport by electrons with $\epsilon$ above the threshold $\epsilon_{\rm th}(T)$. In contrast to the conventional mobility edge, the transport properties of these electrons are not describable in terms of a Drude-like theory, possibly with small weak-localization corrections, no matter how large $\epsilon$ is. Here, although conducting, electron states with arbitrarily large $\epsilon$ remain very much different from the plane waves damped by uncorrelated disorder-induced scatterings. Electron dynamics is, in fact, of the type termed in Ref.~\cite{gornyi05} ``power-law hopping," with hops between quasilocalized states, mediated by electron-electron interactions. This is because the characteristic golden-rule scattering rate (inverse lifetime of the quasilocalized states) $1/\tau_{hc}$ is much smaller than the single-particle spacing within the localization volume for hot electrons,
\be
\frac{1}{\tau_{hc}}\sim\frac{V_{hc}^2}{\Delta_{3hc}}\ll\Delta_h~,
\label{16}
\ee
at the threshold $\epsilon=\epsilon_{\rm th}(T)$ and, for that matter, for arbitrarily large $\epsilon$ above it. The inequality (\ref{16}) means that single-particle levels with energies above $\epsilon_{\rm th}(T)$, despite being sufficiently broadened to ensure many-body delocalization, are still well-resolved in the two-particle correlations that determine transport.

\section{Two dimensions}
\label{s4}

Having established the general framework to describe many-body delocalization in the limit of low $T$ by example of one-dimensional systems in Sec.~\ref{s3}, we now turn to the case of two dimensions. The main difference, compared to one dimension, is that $\xi(\epsilon)$ for two dimensions, in the limit of weak disorder, is an exponentially fast function of $\epsilon$ \cite{lee85}:
\be
\ln\frac{\xi(\epsilon)}{v\tau(\epsilon)}\simeq \pi\epsilon\tau(\epsilon)
\label{17}
\ee
(in the absence of a magnetic field, see below for a generalization), where $1/\tau(\epsilon)$ is the ``transport scattering rate" [which determines the diffusion coefficient at energy $\epsilon$, given by $v^2\tau(\epsilon)/2$]. As seen from Eq.~(\ref{17}), $\xi(\epsilon)$ in two dimensions is a strong function of $\epsilon$ irrespective of the particular form of the correlator of a random potential, or the particular form of the dispersion relation. Specifically, for a quadratic dispersion relation, the density of states (per spin) $\rho=m/2\pi$ is independent of $\epsilon$, so that $\tau$ for the case of short-range disorder with the correlator of the random potential $\langle U({\bf r})U(0)\rangle=w_2\delta ({\bf r})$ does not depend on $\epsilon$, either:
\be
1/\tau=mw_2~.
\label{18}
\ee
As a result, in the limit of short-range disorder, the right-hand side of Eq.~(\ref{17}) is exactly a linear function of $\epsilon$, for arbitrary $\epsilon$. However, no matter whether $\tau$ is exactly ${\rm const}(\epsilon)$ or---as is the case for smooth disorder in two dimensions---a power-law function of $\epsilon$, the characteristic scale of energy on which $\xi(\epsilon)$ changes with varying $\epsilon$ is given by $1/\tau(\epsilon)\ll\epsilon$. Let us focus, for definiteness, on the limit of short-range disorder [Eq.~(\ref{18})].

As before, we introduce, in precisely the same manner, the $h$- and $c$-subsystems and consider the limit of low $T\ll\Delta_c=1/\rho\xi_c^2$. Proceeding along the lines of Sec.~\ref{s3}, we obtain
\be
n_T\sim\frac{l_M}{\xi_c^3}\left(\frac{T}{\Delta_c}\right)^2
\label{19}
\ee
for the spatial density of thermal excitation in the $c$-system. Note the emergence of the Mott length $l_M$ [Eq.~(\ref{6})] in $n_T$ [this is in contrast to Eq.~(\ref{5}): for arbitrary $D$, the power of $l_M$ in $n_T$ can be calculated as $D-1$, which follows from the fact that resonant transitions typically occur over the distance $l_M$ with a small tolerance of the order of $\xi_c$]. Similarly to Eq.~(\ref{8}), the characteristic three-particle spacing of final states for the decay of an $h$-state into another $h$-state and two $c$-states is written as
\be
\Delta_{3hc}=\frac{\Delta_h}{n_T\xi_h^2}~.
\label{20}
\ee

For the characteristic amplitude of the matrix element for coupling between the $h$- and $c$-systems, we have
\be
V_{hc}\sim\alpha T\,\frac{\xi_cl_M}{\xi_h^2}~,
\label{21}
\ee
similar to Eq.~(\ref{11}) for one dimension. The coupling constant $\alpha$ is now given by the two-dimensional integral
\be
\alpha=\frac{m}{2\pi}\left|\int\!d{\bf r}\,V({\bf r})\right|~.
\label{22}
\ee
Note that $V_{hc}$ scales with $\xi_h$ as $1/\xi_h^2$, less rapidly than $\Delta_{3hc}\propto 1/\xi_h^4$ from Eq.~(\ref{20}). It follows that the parameter $\eta_{hc}$ [Eq.~(\ref{9})] can be made arbitrarily large by increasing $\epsilon$. In terms of the single-particle spacings in the $h$- and $c$-systems, $\eta_{hc}$ is represented as
\be
\eta_{hc}\sim\alpha\,\frac{\Delta_c}{\Delta_h}\!\left(\frac{T}{\Delta_c}\right)^3\ln^2\frac{\Delta_c}{T}~.
\label{23}
\ee
When written in this form, the only difference between $\eta_{hc}$ for one and two dimensions [cf.\ Eq.~(\ref{13})] is in the power of the logarithm (which is given by $D$). The essential difference, however, is that now the ratio
\be
\frac{\Delta_c}{\Delta_h}=\frac{\xi_h^2}{\xi_c^2}\simeq\frac{\epsilon}{\epsilon_F}\,\exp\, [\,2\pi(\epsilon-\epsilon_F)\tau\,]
\label{24}
\ee
and, therefore, $\eta_{hc}$ is an exponential of $\epsilon$.

For the threshold energy $\epsilon_{\rm th}(T)$ we thus obtain only a logarithmic divergency with decreasing $T$:
\be
\epsilon_{\rm th}(T)\simeq\epsilon_F+\frac{1}{2\pi\tau}\ln\left[\,\frac{1}{\alpha}\left(\frac{\Delta_c}{T}\right)^3\right]
\label{25}
\ee
(written with a logarithmic accuracy). The conductivity $\sigma(T)$ follows, then, the Arrhenius law with an activation gap that is proportional to $1/\tau$ and weakly (logarithmically) grows as $T$ is decreased, namely [cf.\ Eq.~(\ref{15})]
\be
\ln\sigma(T)\simeq -\frac{1}{2\pi T\tau}\ln\left[\,\frac{1}{\alpha}\left(\frac{\Delta_c}{T}\right)^3\right]~.
\label{26}
\ee
Most importantly, the system remains conducting down to arbitrarily small but finite $T$.

In Eq.~(\ref{17}) for $\xi(\epsilon)$, we assumed that the system is time-reversal invariant. Applying a magnetic field may affect the localization properties of a weakly-disordered two-dimensional system in a substantial way. In particular, in an intermediate range of the magnetic field, defined by the condition that time-reversal symmetry is already totally broken on the spatial scale of the mean free path but the quantum Hall effect does not yet set in, $\xi(\epsilon)$ is given by \cite{hikami81}
\be
\ln\frac{\xi(\epsilon)}{v\tau(\epsilon)}\simeq [\,\pi\epsilon\tau(\epsilon)\,]^2~.
\label{27}
\ee
A straightforward generalization to this case only replaces the logarithmic factor in Eqs.~(\ref{25}) and (\ref{26}) by its square root.

\section{Summary}
\label{s5}

We have discussed many-body localization in a continuum (no lattice), where the single-particle localization length $\xi(\epsilon)$ grows without bound with increasing electron energy $\epsilon$. Many-body localization has been shown not to survive the unbounded growth of $\xi(\epsilon)$, irrespective of the dimensionality of the system. Within our approach, the system in a continuum is characterized by an effective single-particle mobility edge which depends on the temperature $T$ and diverges as $T$ decreases. One of the remarkable consequences of that is the superactivated behavior of the conductivity which vanishes with decreasing $T$ faster than in the Arrhenius law. Delocalization is mediated by interactions between hot electrons with energies above the effective mobility edge
and cold electrons in the vicinity of the Fermi level. Interactions among electrons of only one type play essentially no role.

The mechanism of delocalization studied in the present paper can be generalized to the hybridization channels that involve excitations consisting of {\it multiple} hot electrons. One of them is spectral diffusion \cite{gornyi16a} which might potentially increase the strongly suppressed conductivity obtained here. Another one is coupling between rare hot ``fireballs" \cite{ros15,deroeck16}. Specifically, rare configurations with an anomalously high local energy density were argued to be resonantly coupled with each other and form a delocalized network \cite{ros15,deroeck16}. The hybridization mechanism behind this scenario has a certain analogy with the mechanism studied here: in both cases, the relevant localization length grows with increasing total energy of the excitation. The key difference, however, is in the nature of hot excitations (single-particle vs multiparticle). The simplest of the hot multiparticle excitations are pairs of interacting particles. These were indeed shown to have a larger, for sufficiently strong interactions, localization length compared to single-particle excitations \cite{dorokhov90,shepelyansky94,imry95,vonoppen96}. Within this perspective, it would be interesting to explore delocalization mechanisms in lattice systems, where the single-particle localization length remains bounded, in the spirit of the approach presented in this paper.

Another possibility to enhance delocalization might be provided by more complex couplings between cold states, mediated by polarization of hot electrons, similarly to the mechanism proposed in Ref.~\cite{nandkishore14a}. The contribution of highly excited electrons with a large localization length to the dynamically screened interaction, albeit thermally suppressed, is expected to be of a long-range character. If the power-law decay of random interactions is sufficiently slow, a resonant network appears, independently of the interaction strength \cite{burin98,yao14,gutman16}. It remains to be seen whether this hybridization mechanism gives many-body delocalization and competes, in this sense, with the mechanism studied in the present paper.

A problem closely related to the one studied here concerns many-body localization for the case of a single-particle spectrum in which there is a critical energy $\epsilon_c$ (or a countable set of these) at which the localization length diverges as $\xi(\epsilon\to\epsilon_c)\propto |\epsilon-\epsilon_c|^{-\nu}$. This question was addressed in Ref.~\onlinecite{nandkishore14a} by studying the hybridization processes mediated by resonant couplings of higher order in the interaction strength, with the conclusion that many-body localization is unstable unless the critical exponent $\nu$ is smaller than the bound given by the Harris criterion. The lowest-order cold-bath mechanism of delocalization, studied in the present paper, and its possible generalizations mentioned above could yield a different delocalization criterion. We relegate this study to future work.

\acknowledgments

This work was supported by Russian Science Foundation under Grant No.\ 14-42-00044 (I.V.G.\ and A.D.M.).


\begin{thebibliography}{99}

\bibitem{anderson58} P.W. Anderson, Phys.\ Rev.\ {\bf 109}, 1492 (1958).

\bibitem{fleishman80}  L. Fleishman and P.W. Anderson, Phys.\ Rev.\ B {\bf 21}, 2366 (1980).

\bibitem{gornyi05} I.V. Gornyi, A.D. Mirlin, and D.G. Polyakov,  Phys.\ Rev.\ Lett.\ {\bf 95}, 206603 (2005).

\bibitem{basko06} D.M. Basko, I.L. Aleiner, and B.L. Altshuler,  Ann.\ Phys.\ (N.Y.) {\bf 321}, 1126 (2006).

\bibitem{nandkishore15} R. Nandkishore and D.A. Huse, Annu.\ Rev.\ Condens.\ Matter Phys.\ {\bf 6}, 15 (2015).

\bibitem{imbrie16} J.Z. Imbrie, J.\ Stat.\ Phys.\ {\bf 163}, 998 (2016).

\bibitem{oganesyan07} V. Oganesyan and D.A. Huse, Phys.\ Rev.\ B {\bf 75}, 155111 (2007).

\bibitem{altshuler97} B.L. Altshuler, Y. Gefen, A. Kamenev, and L.S. Levitov, Phys.\ Rev.\ Lett.\ {\bf 78}, 2803 (1997).

\bibitem{ros15} V. Ros, M. M{\"u}ller, and A. Scardicchio, Nucl.\ Phys.\ B {\bf 891}, 420 (2015); {\bf 900}, 446 (2015).

\bibitem{gornyi16} I.V. Gornyi, A.D. Mirlin, and D.G. Polyakov, Phys.\ Rev.\ B {\bf 93}, 125419 (2016).

\bibitem{gornyi16a} I.V. Gornyi, A.D. Mirlin, D.G. Polyakov, and A.L. Burin, arXiv:1611.02681.

\bibitem{deroeck16} W. de Roeck, F. Huveneers, M. M\"uller, and M. Schiulaz, Phys.\ Rev.\ B {\bf 93}, 014203 (2016).

\bibitem{kagan85} Yu. Kagan and L.A. Maksimov, Sov.\ Phys.\ JETP {\bf 61}, 583 (1985).

\bibitem{nandkishore14} R. Nandkishore, Phys.\ Rev.\ B {\bf 90}, 184204 (2014).

\bibitem{mueller09} M. M\"uller, Ann.\ Phys.\ (Berlin) {\bf 18}, 849 (2009).

\bibitem{nandkishore14a} R. Nandkishore and A.C. Potter, Phys.\ Rev.\ B {\bf 90}, 195115 (2014).

\bibitem{abouchacra73} R. Abou-Chacra, D.J. Thouless, and P.W. Anderson, J. Phys. C {\bf 6}, 1734 (1973).

\bibitem{gorkov83} L.P. Gor'kov, O.N. Dorokhov, and F.V. Prigara, Sov.\ Phys.\ JETP {\bf 58}, 852 (1983).

\bibitem{mott70} N.F. Mott, Philos.\ Mag.\ {\bf 17}, 1259 (1968); {\bf 22}, 7 (1970).

\bibitem{lee85} P.A. Lee and T.V. Ramakrishnan, Rev.\ Mod.\ Phys.\ {\bf 57}, 287 (1985).

\bibitem{hikami81} S. Hikami, Phys.\ Rev.\ B {\bf 24}, 2671 (1981).

\bibitem{dorokhov90} O.N. Dorokhov, Sov.\ Phys.\ JETP {\bf 71}, 360 (1990).

\bibitem{shepelyansky94} D.L. Shepelyansky, Phys.\ Rev.\ Lett.\ {\bf 73}, 2807 (1994).

\bibitem{imry95} Y. Imry, Europhys.\ Lett.\ {\bf 30}, 405 (1995).

\bibitem{vonoppen96} F. von Oppen, T. Wettig, and J. M{\"u}ller, Phys.\ Rev.\ Lett.\ {\bf 76}, 491 (1996).

\bibitem{burin98} A.L. Burin, D. Natelson, D.D. Osheroff, and Y. Kagan, in \textit{Tunneling Systems in Amorphous and Crystalline Solids}, edited by  P.\ Esquinazi (Springer, Berlin, 1998).

\bibitem{yao14} N.Y. Yao, C.R. Laumann, S. Gopalakrishnan, M. Knap, M. M{\"u}ller, E.A. Demler, and M.D. Lukin, Phys.\ Rev.\ Lett.\ {\bf 113}, 243002 (2014).

\bibitem{gutman16} D.B. Gutman, I.V. Protopopov, A.L. Burin, I.V. Gornyi, R.A. Santos, and A.D. Mirlin, Phys.\ Rev.\ B {\bf 93}, 245427 (2016).

\end{thebibliography}
\end{document}